# On ductile-regime elliptical vibration cutting of silicon with identifying the lower bound of practical nominal cutting velocity

Jianjian Wang[1, 2, 3], Yang Yang[2], Zhiwei Zhu[4], Yaoke Wang[1], Wei-Hsin Liao[2, 3], Ping Guo[1*]

1. Department of Mechanical Engineering, Northwestern University, Evanston, IL, USA
2. Department of Mechanical and Automation Engineering, The Chinese University of Hong Kong, China
3. Shun Hing Institute of Advanced Engineering, The Chinese University of Hong Kong, China
4. School of Mechanical Engineering, Nanjing University of Science and Technology, Nanjing, China

**Abstract**

   This study investigates the ductile-to-brittle transition behavior in elliptical vibration cutting (EVC) of silicon and identifies the practical process window for ductile-regime cutting. EVC has been reported to increase the critical depth of ductile-regime cutting of silicon. This study demonstrates that the enhanced ductile cutting performance, however, is only optimal in a carefully-determined process window. The vibration amplitudes and nominal cutting velocity have significant impacts on the ductile-to-brittle transition behaviors. Systematic experiments covering a wide span of vibration amplitudes and cutting velocity have been conducted to investigate their effects. Two quantitative performance indices, the critical depth and ductile ratio, are utilized to analyze cutting performance by considering two unique characteristics of elliptical vibration cutting, i.e., the time-varying undeformed chip thickness and effective cutting direction angle. The results show that there exists a lower bound for the nominal cutting velocity to ensure the ductile-regime material removal, besides the well-known upper bound. Besides, the increases of vibration amplitudes in both the cutting and depth-of-cut (DOC) directions first enhance but then deteriorate the cutting performance. Based on the theoretical analysis and experimental results, the optimal process parameters have been recommended for the elliptical vibration cutting of silicon.

∗ Corresponding author: Ping Guo. Email: ping.guo@northwestern.edu
   First author: Jianjian Wang. Email: jjwang@northwestern.edu

## Nomenclature:

| | | | |
|---|---|---|---|
| DOC | Depth of cut | $R$ | Tool nose radius |
| $x$ | Displacement in the cutting direction | $\gamma_0$ | Tool rake angle |
| $y$ | Displacement in the DOC direction | $r$ | Tool cutting edge radius |
| $a$ | Vibration amplitude in the cutting direction | $w$ | Cutting width |
| $b$ | Vibration amplitude in the DOC direction | $a_{p,\text{critical}}$ | Critical depth |
| $v_c$ | Nominal cutting velocity | $\alpha_{\text{ductile}}$ | Ductile ratio |
| $f$ | Vibration frequency | $A_{\text{ductile}}$ | Ductile removal area |
| $a_p$ | Nominal DOC | $A_{\text{total}}$ | Total removal area |
| $t$ | Time | $\theta_a$ | Average cutting direction angle |
| $h_c$ | Minimum chip thickness | $\theta$ | Instantaneous cutting direction angle |
| $h_{\max}$ | Maximum chip thickness | $\theta_c$ | Critical cutting angle |
| $h_{\text{ductile}}$ | Upper critical chip thickness | $\gamma$ | Effective rake angle |

## 1. Introduction

Silicon, as a brittle material with intrinsic low fracture toughness, is difficult for mechanical machining to generate a crack-free surface. Fang and Venkatesh (1998) reported that with the increase of undeformed chip thickness, the material removal mechanism of silicon undergoes a transition from the ductile regime to the brittle regime. Fang et al. (2007) demonstrated from machining tests on silicon that, in the ductile regime, silicon is removed through plastic deformation due to phase transformation to obtain a crack-free surface; while in the brittle regime, the silicon surface severely fractures due to the crack propagation. Traditionally, the finishing of silicon surfaces mainly replies on polishing to remove subsurface cracks induced by preceding machining processes such as grinding and lapping. Alternatively, ultra-precision machining using a single point diamond tool has been proposed to direct machine silicon in ductile-regime at a very fine scale to achieve nanometric surface finish due to the extreme sharpness of the tool. Abdulkadir et al. (2018) performed a critical review on ductile-regime cutting of silicon. They suggested that the ultra-precision diamond cutting eliminates the multi-step machining steps in the grinding/lapping and polishing combinations and provides unique advantages in terms of stringent form accuracy and high flexibility. However, owing to the small ductile-to-brittle transition distance, the depth-of-cut (DOC) in single-point diamond cutting of silicon is strictly limited. This critical DOC separates the cutting condition from the regime of plastic deformation to the brittle fracture removal, which restricts the machining efficiency and process flexibility.

To improve the critical depth in ductile-regime cutting of brittle materials, Suzuki et al. (2004)

proposed to use elliptical vibration cutting to suppress the crack nucleation. Moreover, Huang et al. (2018) explained that, as shown in Fig. 1, the distance from cutting path to the target surface in elliptical vibration cutting can tolerate some degrees of crack propagation. Elliptical vibration cutting has demonstrated some promising results in ductile-regime machining of brittle materials with improved critical depth. For example, Geng et al. (2020) utilized elliptical vibration to reduce the cutting temperature and improve surface integrity in the machining of brittle materials. Zhu et al. (2016) used elliptical vibration cutting to increase the critical depth of silicon from 58 nm to 745 nm compared with conventional diamond cutting. Zhang et al. (2017) applied elliptical vibration cutting for micro-structured surface texturing of silicon with the improvement ratio of critical depth of up to 12.5 times.

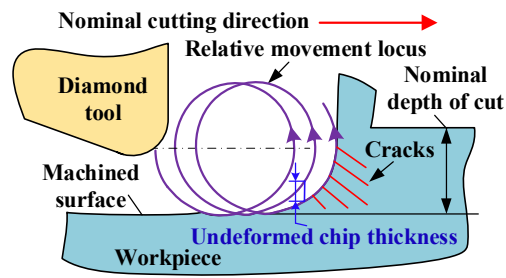

**Fig. 1.** Illustration of elliptical vibration cutting of brittle materials.

In elliptical vibration cutting, the process parameters, including the nominal cutting velocity and vibration amplitudes deterministically affect the ductile-to-brittle transition behavior of brittle materials such as silicon, and ultimately define the quality of machined surfaces. Suzuki et al. (2004) experimentally studied the elliptical vibration cutting performance of tungsten carbide and zirconia ceramics. Their results showed that elliptical vibration cutting could improve the surface quality only if the vibration amplitude in the DOC direction is comparable to the size of material grains. As a follow-up, Zhang et al. (2014) also conducted elliptical vibration cutting experiments on tungsten carbide to explore the effects of elliptical amplitudes on the critical depth. They demonstrated that the vibration amplitudes in both the cutting and DOC directions significantly affect the surface quality of tungsten carbide processed by vibration cutting. Moreover, Nath et al. (2011) investigated the effects of nominal cutting velocity on the ductile-regime cutting performance. The results indicated that the nominal cutting velocity must be kept below a critical value to improve the critical depth, which is known as the upper bound of practical nominal cutting velocity to ensure the performance of elliptical vibration cutting.

Several attempts have been made to explain the deterministic effects of process parameters on the ductile-to-brittle transition behavior in elliptical vibration cutting of brittle materials. Nath et al. (2011) developed a model by analyzing the undeformed chip thickness to calculate the critical nominal cutting velocity. If the nominal cutting velocity exceeds the critical value, there is a sharp increase of undeformed chip thickness within a vibration cycle. Huang et al. (2018) also studied the effects of nominal cutting velocity and vibration amplitudes on the critical depth by analyzing undeformed chip thickness. Their model demonstrated that the critical depth increases with the cutting-direction vibration amplitude and the nominal cutting velocity. Besides, Zhang et al. (2019) utilized a criterion based on the specific cutting energy to model the critical depth for elliptical vibration cutting of reaction-bonded silicon carbide. The above three theoretical studies based on the analysis of undeformed chip thickness partially explained the effects of process parameters on ductile-regime cutting of silicon; however, the experimental verifications were not comprehensive enough to account for the interactive effects of process parameters. Moreover, Dai et al. (2019) utilized molecular dynamics simulation to understand the effects of elliptical vibration on the material behavior of silicon in nano-cutting. The applicable DOC in the simulation, however, was limited by the computing power as small as 1.5 nm, which is far from the microscale DOC in practice.

The existing knowledge regarding the ductile-to-brittle transition behavior in elliptical vibration cutting of brittle materials, especially silicon, is still very limited, which leads to some inconsistent and contradictory results regarding its ductile-regime cutting performance. As reported by Wang et al. (2018), silicon surfaces machined by elliptical vibration cutting were totally fractured, even the adopted nominal cutting velocity was kept smaller than its critical value. As this study is to show, which has not been reported before, the nominal cutting velocity also has a lower bound below which the machined surface will be severely fractured. One possible reason for the lack of full understanding is the scattered experimental design in available studies, failing to cover a wide range of process parameters in elliptical vibration cutting. Besides, most research has been focused only on the analysis of undeformed chip thickness. Another essential process characteristic, namely the time-varying effective cutting direction, has not been well considered. The time-varying cutting direction correlates to the effective rake angle, which affects the stress state at the cutting zone. Its influence on the ductile-to-brittle transition behavior should not be ignored.

Moreover, as demonstrated by Gu et al. (2018), most studies only utilized a onefold performance index, i.e., the critical depth to characterize the grooved surface to explore the process effects on the ductile-to-brittle transition of silicon. However, the critical depth alone cannot adequately describe the material removal state in the more complicated case for elliptical vibration cutting. Yuan et al. (2015) utilized a ductile-to-brittle ratio to characterize the material removal state in ultrasonic grinding of ceramics composites. This method is adopted in this study to add a quantitative metric for the evaluation of machined surface quality.

This study presents the surface quality evaluation in the elliptical vibration cutting of silicon, particularly in an attempt to answer the question of how the vibration amplitudes and nominal cutting velocity affect the ductile-to-brittle transition behavior. The experimental design covers a wide span of process parameters, up to one hundred sets of grooving tests with distinct combinations of different elliptical vibration trajectories and nominal cutting velocities. The critical depth and the ductile ratio of machined surfaces are used as the metrics to quantitatively study the material transition behavior in elliptical vibration cutting of silicon. Besides, both the time-varying undeformed chip thickness and the effective cutting direction are evaluated. The optimal process parameters to achieve the best ductile cutting performance are recommended based on our findings.

## 2. Experimental design

Inclined grooving tests are used in this study to investigate the ductile-to-brittle transition characteristics in silicon cutting under different vibration trajectories and nominal cutting velocities. The experiments are systematically designed to cover a wide span of process parameters (up to 100 sets of experiments) due to the highly complicated behavior in the ductile-regime cutting, as analyzed above. The critical depth and the ductile ratio are adopted as the characteristic performance indices to identify the process window for elliptical vibration cutting of silicon in ductile-regime.

## 2.1 Definition of performance indices: critical depth and ductile ratio

Two quantitative indices, the critical depth and ductile ratio, are adopted to describe the characteristics of machined surfaces, considering the complexity of material removal states in elliptical vibration cutting. Fig. 2(a) and 2(b) demonstrate a representative optical image of the machined silicon groove and its cutout image, respectively. To identify the material removal state from the surface morphology, the binarization method has been utilized to process the optical image by setting a grayscale threshold at 100. The resulting binary image is shown in Fig. 2(c).

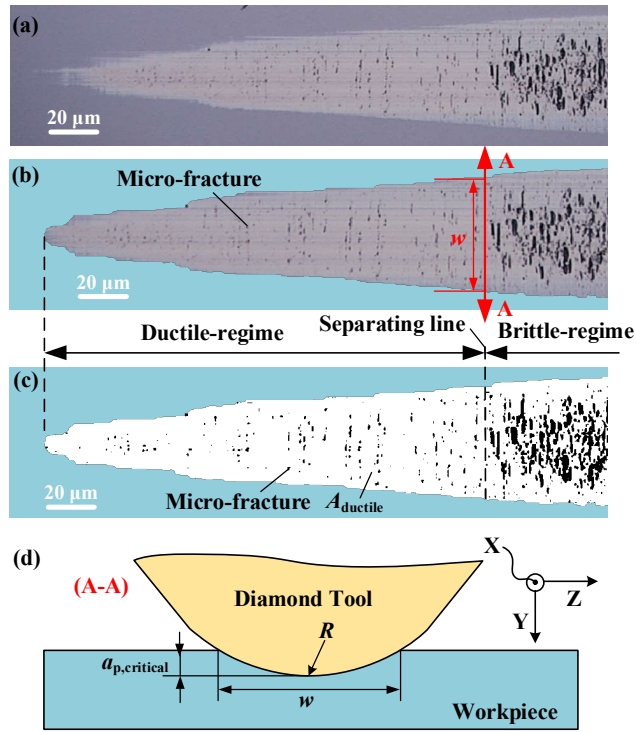

**Fig. 2.** Evaluation method of machined groove surfaces: (a) original optical image; (b) cutout image; (c) binary image of the silicon groove; and (d) calculation of the critical depth.

### 2.1.1 Critical depth

The critical depth $a_{p,critical}$ separates the ductile and brittle cutting regimes, which is indicated by the sudden deterioration of surface integrity. As shown in Fig. 2(b) and 2(c), the gradual increase of width $w$ of the machined groove from left to right indicates the increase of cutting depth. The surface quality of the machined groove deteriorates sharply with dense and bulky surface cracks arising when the cutting depth exceeds a critical value as marked by (A-A) in Fig. 2(b), where (A-A) is the line separating the ductile and brittle cutting regimes. According to the cross-sectional view of the groove along (A-A) as shown in Fig. 2(d), the critical depth $a_{p,critical}$ can be calculated from the groove width $w$ and nose radius $R$ of the diamond tool by

$$a_{p,critical} = R - \sqrt{R^2 - (w/2)^2} \tag{1}$$

*2.1.2 Ductile ratio*

Another quantitative index used to represent the surface quality is the ductile ratio. It is defined as the ratio between the crack-free surface area and the machined area. The ductile ratio termed as $\alpha_{ductile}$ is calculated by

$$\alpha_{ductile} = \frac{A_{ductile}}{A_{total}} \tag{2}$$

where $A_{ductile}$ and $A_{total}$ are the ductile removal area and total area, respectively, as shown in Fig. 2(c). $A_{ductile}$ is determined from the white pixel counts in the binary image, while the total area $A_{total}$ is the summation of the white and black pixels.

The ductile ratio can describe the quality of machined surfaces quantitatively. A visualized relationship between the ductile ratio and surface morphology is given in Fig. 3, which demonstrates the cases of different calculated ductile ratios. When the ductile ratio is at 94%, the surface quality is considered unacceptable due to the dense crack distribution. We here set the ductile ratio of 97% as a threshold value of data selection when plotting the figure regarding the quantitative effects of elliptical vibration on the critical depth $a_{p,critical}$.

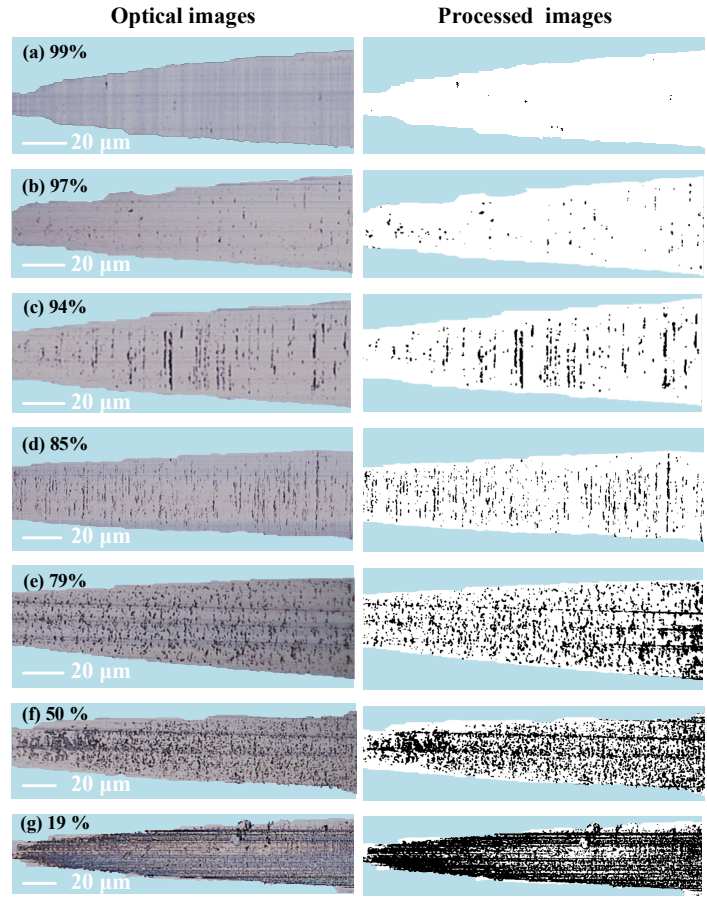

**Fig. 3.** Examples of visual correlation between the surface morphology and ductile ratio.

### *2.1.3 Combined utilization of critical depth and ductile ratio*

The critical depth and ductile ratio should be utilized together to characterize the ductile-to-brittle transition of machined surfaces in elliptical vibration cutting of brittle materials. The material removal state in conventional diamond cutting of brittle materials is regarded to transit from perfect ductile-regime (ductile ratio = 100%) to brittle-regime with the increase of cutting depth. In this case, only the critical depth is sufficient to characterize the ductile-to-brittle transition of machined surfaces. However, as this study is to show, the material removal state of machined surfaces in elliptical vibration cutting becomes more complicated. On the one hand, an apparent ductile-to-brittle transition of material removal state can be identified from the sharp variation of machined surface quality with the increase of cutting depth. On the other hand, the machined surface morphology is often not in perfect ductile-regime even if the cutting depth is minimal. Hence, in this study, the critical depth and ductile ratio are utilized synergistically to evaluate the ductile-to-brittle transition of machined surfaces.

### 2.2 Experimental setup for grooving tests

The experimental schematic and platform are described as follows. Fig. 4(a) shows the principle of inclined groove cutting, where a 2-D ultrafast vibration cutting tool is used to provide prescribed elliptical trajectories. The ultrafast tool is designed by Wang et al. (2019) to have a frequency bandwidth of up to 6 kHz and a workspace of 16 μm × 10 μm to generate high-frequency elliptical vibration in a fully controllable manner. The workpiece, pre-polished silicon wafer with the surface roughness better than 0.5 nm, has a crystal plane of (100). It is tilted at an angle of 0.002 radians (2 μm/mm) to introduce a gradual increase of DOC. All cuttings are performed along the crystal orientation of [110].

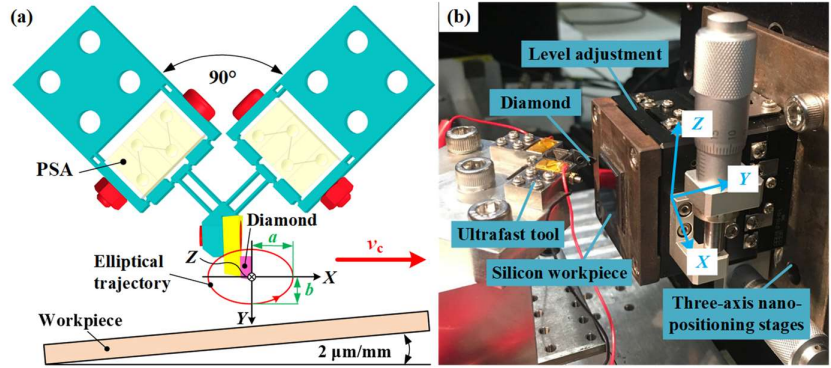

**Fig. 4. Description of experimental setup: (a) schematic of groove cutting and (b) hardware platform.**

Fig. 4(b) shows a detailed view of the actual setup for grooving tests. A single-crystal diamond insert with a nose radius $R$ of 470 μm, rake angle $\gamma_0$ of -20° and clearance angle of 10°, is mounted on the ultrafast tool. The tool is stationed on a linear driver (ACT165DL, *Aerotech, Pittsburgh, USA*), which is used to generate the nominal cutting motion. The silicon workpiece is glued on a two-degree-of-freedom leveling stage, which adjusts the slope angle of the workpiece to 0.002 radians. The leveling stage is mounted on a three-axis positioning platform (ANT165XY and ANT165LZ, *Aerotech, Pittsburgh, USA*), which sets the nominal DOC and cross-feed.

## 2.3 Experimental design for grooving tests

Five groups of grooving tests were performed with all the combinations of process parameters summarized in Table. 1. First, the grooving tests with conventional diamond cutting without elliptical vibration were performed to set the baseline performance. The second experimental group was designed to identify the effects of nominal cutting velocity by using a set of fixed elliptical trajectories (vibration amplitude in cutting direction $a$=3 μm, vibration amplitude in DOC direction $b$=0.5, 1, or 2 μm) with thirteen different nominal cutting velocity ranging from 0.02 mm/s to 3 mm/s. The last three groups were designed to investigate the effects of the elliptical trajectory shape. The vibration amplitude in the cutting direction was set from 0.5 μm to 4 μm with an increment interval of 0.5 μm, while the vibration amplitude in the DOC direction was set at 0.5, 1, and 2 μm. In addition, each group of tests utilized a different nominal cutting velocity at 0.1, 0.2, and 0.4 mm/s respectively to analyze the interactive effects of cutting velocity and vibration amplitudes.

**Table. 1 Summary of the processing parameters in the experimental design.**

| No. | $a$ (μm) | $b$ (μm) | Nominal cutting velocity $v_c$ (mm/s) |
|---|---|---|---|
| 1 | Without vibration | | 0.02, 0.05, 0.1, 0.15, 0.2, 0.25, 0.3, 0.35, 0.4, 0.5, 0.6, 0.8, 3 |
| 2 | 3 | 0.5, 1, 2 | 0.02, 0.05, 0.1, 0.15, 0.2, 0.25, 0.3, 0.35, 0.4, 0.5, 0.6, 0.8, 3 |
| 3 | 0.5, 1, 1.5, 2, 2.5, 3, 3.5, 4 | 0.5, 1, 2 | 0.1 |
| 4 | 0.5, 1, 1.5, 2, 2.5, 3, 3.5, 4 | 0.5, 1, 2 | 0.2 |
| 5 | 0.5, 1, 1.5, 2, 2.5, 3, 3.5, 4 | 0.5, 1, 2 | 0.4 |

**Tool vibration frequency:** $f$ = 2000 Hz

**Tool geometries:** nose radius $R$=470 μm, rake angle $\gamma_0$ =-20°, clearance angle=10°

**Cutting direction:** along the crystal orientation of [110] in the crystal plane of (100)

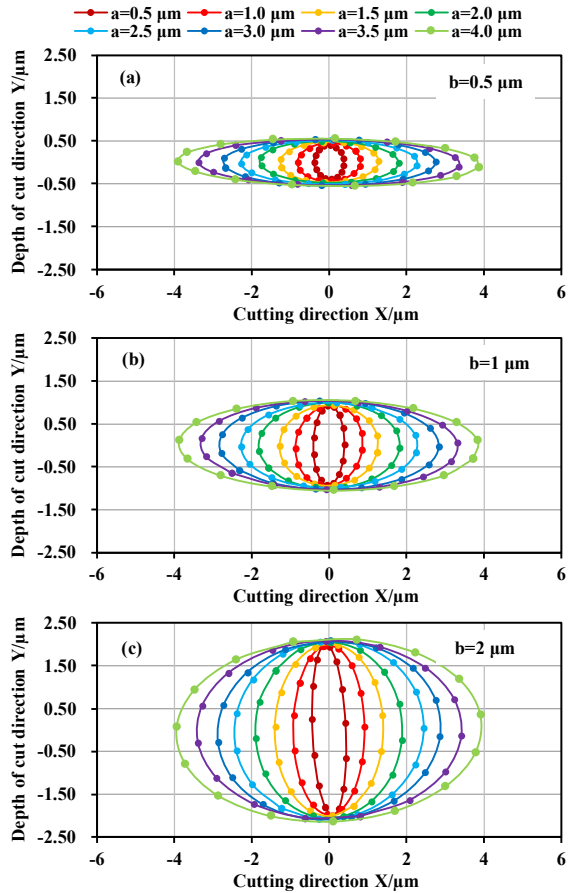

**Fig. 5. Measurement results of designed trajectories for (a) $b$=0.5 μm, (b) $b$=1 μm and (c) $b$=2 μm.**

The tool vibration frequency was kept at 2000 Hz for all grooving tests. The actual elliptical trajectories were measured before cutting to ensure the conformity using two capacitance displacement sensors (*MicroSense*, *USA*). The measurement trajectories are plotted in Fig. 5. Each cutting test was repeated twice. The machined surface was then inspected by a digital microscope (*Hirox* RH-2000, *Tokyo, Japan*).

# 3. Analysis of the ductile-to-brittle transition in silicon cutting

There are both similarities and distinctions with regards to the cutting mechanism between conventional diamond machining and elliptical vibration cutting of silicon in the ductile regime. The ductile-to-brittle transition behavior of silicon cutting with elliptical vibration can be analyzed based on the differentiation of these similarities and distinctions. In this section, a theoretical model is established to describe the unique cutting characteristics of elliptical vibration cutting over the conventional cutting. Then, the ductile cutting criteria of silicon with elliptical vibration is derived based on the proposed model and compared to the existing knowledge in conventional cutting.

## 3.1 Ductile cutting mechanism in conventional diamond cutting of silicon

In conventional diamond cutting of silicon, chip thickness plays a critical role in determining the material removal state. It is well known that, in diamond cutting of silicon, the chip thickness should be smaller than an upper critical value $h_{ductile}$ to avoid material fracture. However, it was noted by Yuan et al. (1996) that there also exists a minimum chip thickness $h_c$, below which the effective negative rake angle is too large that ploughing and sliding take place instead of cutting to generate marring damage. Fang and Zhang (2003) experimentally evaluated the minimum chip thickness $h_c$ in silicon cutting. According to their experimental results, $h_c$ is about 0.27~0.45 times of the edge radius $r$. They also demonstrated that $h_c$ is related to the rake angle: a larger negative rake angle leads to a larger $h_c$ due to the increase of compressive stress below the cutting edge. In short, to ensure the ductile-regime conventional cutting of silicon, the chip thickness should be bounded by $h_c$ and $h_{ductile}$.

In addition to the chip thickness, the tool rake angle is another factor that critically determines the material removal state. As illustrated in Fig. 6(a), the rounded cutting edge (due to the finite edge radius) with a radius $r$ introduces a negative effective rake angle $\gamma_0$, which can generate both high hydrostatic pressure and shear stress on materials below the cutting edge. Yan et al. (2009) demonstrated that the high hydrostatic pressure facilitates the phase transformation of silicon to a metallic phase and shields the propagation of cracks to avoid brittle fracture, while the shear stress prompts the plastic deformation of the metallic phase to induce chip flow in the ductile regime. In practice, according to the experimental results from Yan et al. (2001), a moderate negative rake angle (−20°~−50°) is preferred for ductile-regime cutting.

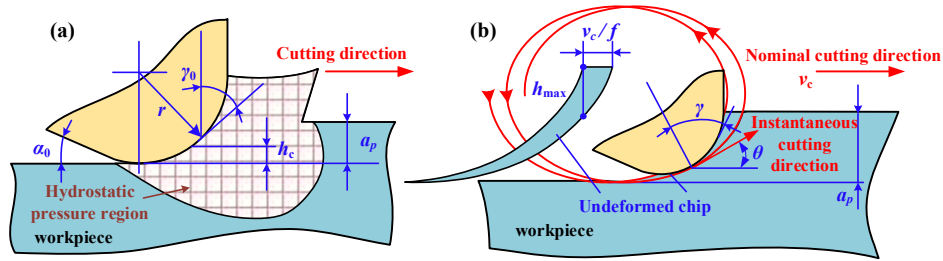

**Fig. 6.** Mechanisms of ductile cutting of brittle materials: (a) conventional cutting; (b) elliptical vibration cutting.

## 3.2 Unique characteristics in elliptical vibration cutting

When it comes to the elliptical vibration cutting of silicon in the ductile regime, the underlying mechanisms become more complicated due to the unique overlapping trajectory of the tool. The moving tool trajectories are determined by the process parameters including the nominal cutting velocity $v_c$, vibration frequency $f$, vibration amplitudes $a$ and $b$ in the cutting and DOC directions as expressed by the following equations,

$$\begin{cases} x = a\sin(2\pi ft) + (2\pi ft) \cdot \dfrac{1}{2\pi} \cdot \dfrac{v_c}{f} \\ y = b\cos(2\pi ft) \end{cases} \quad (3)$$

where $t$ denotes time. As shown in Fig. 6(b) and Eq. (3), it is worth noting that $v_c$ and $f$ act synergistically as $v_c/f$ to determine the moving trajectory of the diamond tool.

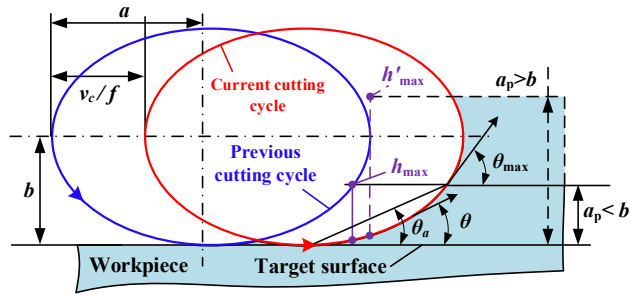

Fig. 7. A simplified model for maximum undeformed chip thickness $h_{max}$ calculation.

The overlapping tool trajectory in elliptical vibration cutting results in two unique characteristics, the time-varying undeformed chip thickness and the time-varying effective cutting direction. Fig. 7 shows a simplified model to illustrate the above two time-varying parameters. When the nominal DOC, $a_p$, is smaller than the vibration amplitude $b$ in the DOC direction, the instantaneous undeformed chip thickness in the current cutting cycle increases from zero to the maximum value $h_{max}$. It then decreases to zero, while the effective cutting direction changes from the horizontal direction to an oblique direction. However, when the DOC, $a_p$, is larger than the vibration amplitude $b$ in the DOC direction, there will be a sudden increase of undeformed chip thickness during the cutting cycle. The process conditions $a_p<b$ can prevent the crack-sensitive impact force induced by the sudden increase of undeformed chip thickness. According to Fig. 7, the maximum undeformed chip thickness $h_{max}$ and the average effective cutting direction denoted by $\theta_a$ can be derived as,

$$\begin{cases} h_{max} = a_p - b + b\sqrt{1 - \left(\sqrt{1 - \left(1 - \dfrac{a_p}{b}\right)^2} - \dfrac{v_c/f}{a}\right)^2} \\ \\ \theta_a = \arctan\left(\dfrac{a_p}{a} \cdot \dfrac{1}{\sqrt{1 - \left(1 - \dfrac{a_p}{b}\right)^2}}\right) \end{cases} \quad (4)$$

Both the maximum undeformed chip thickness $h_{max}$ and effective cutting direction angle $\theta$ have crucial effects on the material transition behavior of silicon cutting. Based on Eq. (4), $h_{max}$ and $\theta$ can be analytically calculated based on the given process parameters. Their relationship with the critical depth $a_{p,critical}$ is then experimentally derived in Fig. 8. These two variables are utilized as the intermediate variables to analyze the interrelated effects of different process parameters (vibration amplitudes, cutting velocity, vibration frequency, etc.) on the ductile-regime cutting performance (critical depth and ductile ratio).

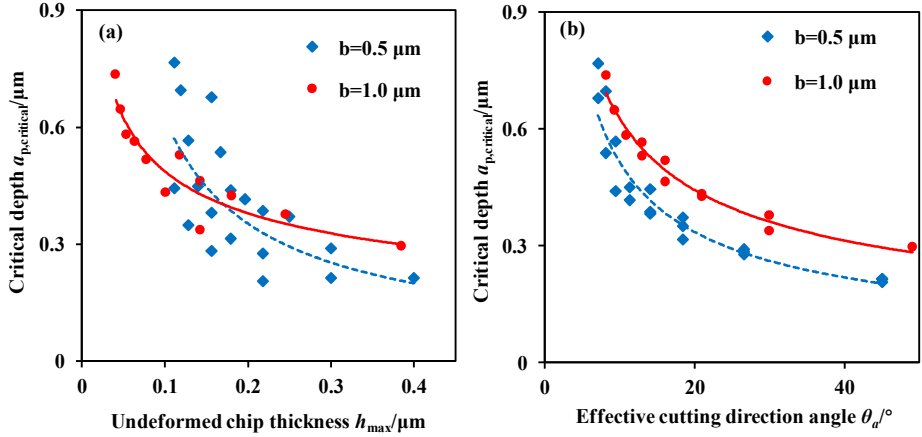

**Fig. 8.** Correlation between the critical depth and (a) undeformed chip thickness and (b) effective cutting direction.

### 3.3 Ductile cutting criteria of silicon with elliptical vibration

Intuitively, the reduction of maximum undeformed chip thickness $h_{max}$ seems helpful to enhance the ductile-regime cutting. On the other hand, the increase of the effective cutting direction ($\theta$) might result in an unfavorable cutting state as shown by the trends in Fig. 8. These two parameters, however, are not independent as indicated by Eq. (4). They will simultaneously affect the cutting characteristics.

First of all, the existence of chip thickness boundary ($h_{ductile}$ and $h_c$) in conventional cutting is still valid in elliptical vibration cutting. During a vibration cycle, the instantaneous undeformed chip thickness will vary with time. The maximum undeformed chip thickness $h_{max}$ is adopted to represent the characteristic length scale in each vibration cycle. It should be bounded by the upper and lower limits, $h_c$ and $h_{ductile}$, as introduced in Section 3.1. The upper bound $h_{ductile}$ ensures the length scale is small enough to favor the ductile removal, while the lover bound $h_c$ ensures the chip formation rather than ploughing.

Secondly, the time-varying effective cutting direction in elliptical vibration cutting changes the effective rake angle $\gamma$, which further affects chip thickness boundary ($h_{ductile}$ and $h_c$). Due to the slope of tool trajectories, as shown in Fig. 6(b), the effective rake angle $\gamma$ is always negative hence helps to increase $h_c$. In the meantime, the negative $\gamma$ might reverse the friction force direction on the tool rake, which would reduce the hydrostatic pressure at the cutting zone and induce crack formation. Zhang et al. (2012) established that the reverse of friction direction takes place when the effective cutting direction angle $\theta$ is larger than $\theta_c=45°+\gamma_0$. The effective cutting direction $\theta$ should always be kept smaller than $\theta_c$, and preferably away from the critical angle $\theta_c$ to keep a positive frictional force on the tool-chip interface.

In summary, the two unique cutting characteristics of elliptical vibration cutting have simultaneous and interactive effects on the ductile-to-brittle transition of silicon. Hence, the improvement of the critical depth is condition-dependent. According to the analysis above, the ductile-regime cutting criteria in terms of the undeformed chip thickness and effective cutting direction can be summarized as,

$$\begin{cases} h_c\left(\gamma = \gamma_0 - \theta\right) < h_{\max} < h_{\text{ductile}}\left(\gamma = \gamma_0 - \theta\right) \\ \gamma = \gamma_0 - \theta > -50° \\ \theta < \theta_c = 45° + \gamma_0 \end{cases} \quad (5)$$

Based on the proposed criteria in Eq. (5) and the definitions of $h_{\max}$ and $\theta$ in Eq. (4), the practical process window can be determined. The intermediate variables, $h_{\max}$ and $\theta$, can be firstly related to the actual process parameters. According to Eq. (4), $h_{\max}$ is dependent on $v_c/f/a$ and $b$, while $\theta$ is dependent on $a$ and $b$. The results are plotted in Fig. 9. At the meantime, according to Eq. (5), to enhance the ductile-regime cutting of brittle materials, $h_{\max}$ should be bounded by $h_c$ and $h_{\text{ductile}}$; while $\theta$ should satisfy the condition $\theta < 25°$ (when $\gamma_0 = -20°$). These criteria are utilized together with the relationship shown in Fig. 9 to analyze the complex interactive effects of process parameters on the ductile-regime cutting performance in the following sections.

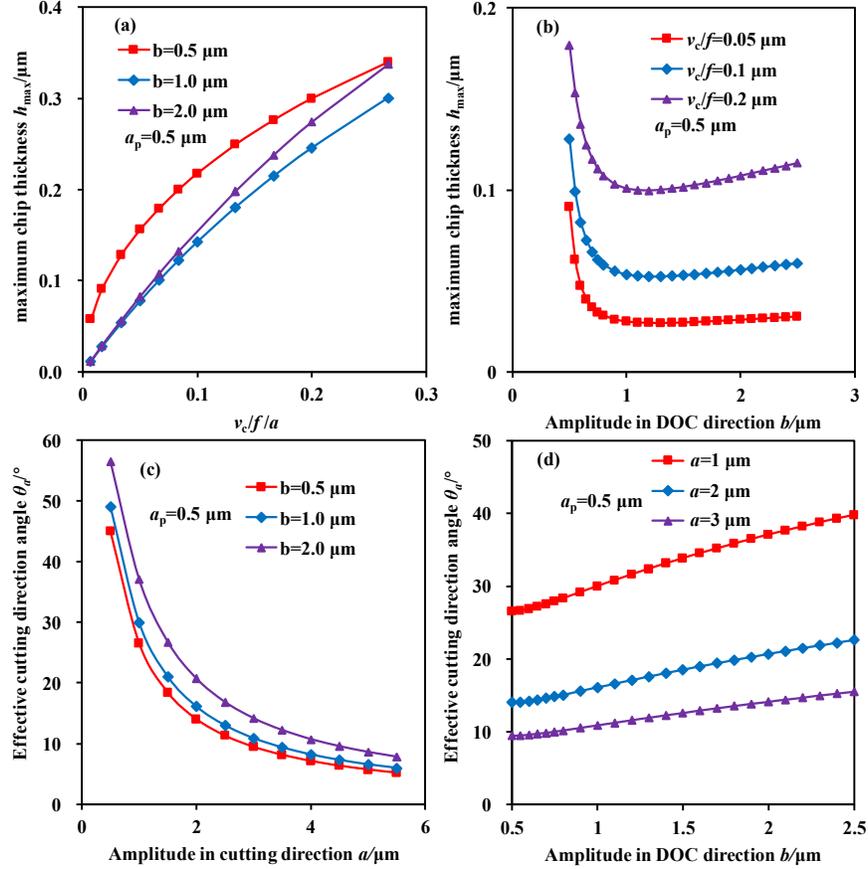

**Fig. 9.** Relationships between the process parameters and the intermedia variables $h_{max}$ (maximum undeformed chip thickness) and $\theta$ (effective cutting direction angle): effects of (a) $v_c/f/a$ and (b) $b$ on $h_{max}$; effects of vibration amplitude (c) $a$ and (d) $b$ on $\theta$.

## 4. Results and discussions

There are four main process parameters that are critical to the ductile-to-brittle transition distance in elliptical vibration cutting of silicon. They are the nominal cutting velocity $v_c$, vibration frequency $f$, vibration amplitude $a$ in the cutting direction, and vibration amplitude $b$ in the DOC direction. These process parameters determine the moving trajectory of the diamond tool, hence influence the variation of undeformed chip thickness and effective cutting direction. Through the focus on the two intermediate variables, i.e., the effective cutting direction and undeformed chip thickness, here we present the experimental results with the theoretical analysis to systematically study the relationship between the ductile-to-brittle transition behavior and the four main process parameters.

### 4.1 Effects of nominal cutting velocity

There exists a lower bound and an upper bound of practical $v_c/f$ to ensure the ductile-regime cutting of silicon. The effects of $v_c/f$ on the ductile cutting performance can be attributed to its direct influence on the undeformed chip thickness. As indicated in Eq. (4), $v_c/f$ is only related to the maximum undeformed chip thickness $h_{max}$ but not to the effective cutting direction $\theta$. The relationship between $v_c/f$ and $h_{max}$ is plotted in Fig. 9(a), while the experimental results with regards to the effects of $v_c/f$ on the critical depth and ductile ratio are illustrated in Fig. 10. As shown in Fig. 9(a), $h_{max}$ monotonically increases with $v_c/f$. Since $h_{max}$ has both an upper and a lower bound to achieve the ductile cutting of silicon ($h_c < h_{max} < h_{ductile}$), it can be derived that there must exist an upper and a lower bound of practical $v_c/f$ for the ductile-regime cutting of silicon.

The existence of the upper bound of practical $v_c/f$ is well-known, which has also been reported by other researchers such as Huang et al. (2018). As shown in Fig. 10(a), $v_c$ did not show significant influences on the critical depth in conventional diamond cutting. In contrast, the critical depth decreased slightly with the increase of nominal cutting velocity in elliptical vibration cutting of silicon. Due to the ductile cutting criteria $h_{max} < h_{ductile}$, with the further increase of $v_c/f$, it will reach the upper bound to be comparable to conventional diamond cutting.

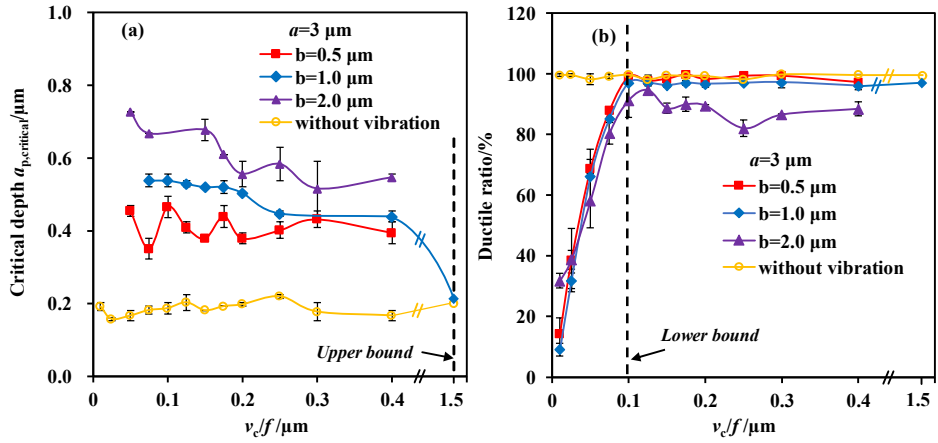

Fig. 10. Effects of nominal cutting velocity/frequency $v_c/f$ on (a) critical depth and (b) ductile ratio.

The existence of the lower bound of $v_c/f$ is clearly demonstrated in Fig. 10(b). When cutting without vibration, though the critical depth is small at around 200 nm, the ductile ratio is up to 99.5% with good surface quality. In elliptical vibration cutting, however, the ductile ratio drops significantly when $v_c/f$ approaches 0, and then maintains around 100% when $v_c/f$ exceeds 0.1 µm. For example, when $v_c/f = 0.01$ µm ($a=3$ µm, $b=0.5$, 1 or 2 µm), the achieved ductile ratio is below 20% showing severe damage as illustrated in Fig. 3(g). Only when $v_c/f$ is greater than 0.1 µm, the elliptical vibration cutting can improve the critical depth by 2~3 times compared to conventional diamond cutting with acceptable surface quality (critical depth: 400~600 nm, while ductile ratio >97%).

The lower bound of $v_c/f$ results from the minimum chip thickness $h_c$. The minimum chip thickness effect has been extensively observed in metal cutting, such as that presented by Son et al. (2005), but is often ignored in single crystal diamond cutting. However, in elliptical vibration cutting of silicon, the effective undeformed chip thickness is comparable to the tool edge radius due to the overlapping trajectories. The minimum chip thickness effect is significant in the ductile-regime cutting, resulting in ploughing and fractured surface. According to the experimental results from Fang and Zhang (2003), the minimum chip thickness $h_c$ in silicon cutting is about 0.27~0.45 times of the edge radius $r$, which is approximately equal to the ductile-regime critical depth of conventional silicon cutting as demonstrated by Liu et al. (2007).

Hence, the estimated $h_c$ is about 54~90 nm, considering the obtained critical depth of 200 nm in this study. When $v_c/f = 0.1$ μm, the maximum undeformed chip thickness $h_{max}$ at the critical depth is calculated as 87 nm ($b$=0.5 μm) and 66 nm ($b$=1 μm), which lies in the estimated range of $h_c$.

**4.2 Effects of vibration amplitude in the depth-of-cut (DOC) direction**

The increase of vibration amplitude $b$ in the DOC direction first improves the ductile-regime cutting of silicon and then deteriorates the cutting performance. The effects of $b$ on the critical depth and ductile ratio are illustrated in Fig. 11 for three different sets of $v_c/f$. The first set of results shown in Fig. 11(a) and Fig. 11(b) are in the ploughing region as analyzed in the last section, which results in the abnormal surface quality. Here we focus on the second and third sets of results shown in Fig. 11(c) to Fig. 11(f) that give a clear and consistent effect of the vibration amplitude $b$.

On the one hand, with the initial increase of vibration amplitude $b$ in the DOC direction, the ductile cutting performance of silicon is improved. As shown in Fig. 11(c) and Fig. 11(d), when $v_c/f = 0.1$ μm and 0.2 μm, the blue curve ($b = 1$ μm) shows consistently larger critical depth compared with the red curve ($b = 0.5$ μm), while both cases have an acceptable ductile ratio (close to 100%). This phenomenon can be directly attributed to the relationship between $b$ and $h_{max}$, as shown in Fig. 9(d). When $b$ increases from 0.5 μm to 1 μm, the maximum undeformed chip thickness $h_{max}$ undergoes a sharp decrease, resulting in the improvement of critical depth.

On the other hand, with the further increase of vibration amplitude $b$ in the DOC direction, the machined surface quality is deteriorated rapidly. As shown in Fig. 11(d) and Fig. 11(f), when $b$ equals to 2 μm, the ductile ratio drops significantly below 90% and becomes even worse (80%) when the cutting velocity ratio $v_c/f$ is further increased. This observation is due to the combined effects of the increase in maximum undeformed chip thickness $h_{max}$ and the effective cutting direction angle $\theta$, as illustrated in Fig. 9(b) and Fig. 9(d). Especially when $b$ increases from 1 μm to 2 μm, the increase in $\theta$ leads to a sudden reverse of friction direction at the interface of tool rake face and chips, causing the severe surface damage.

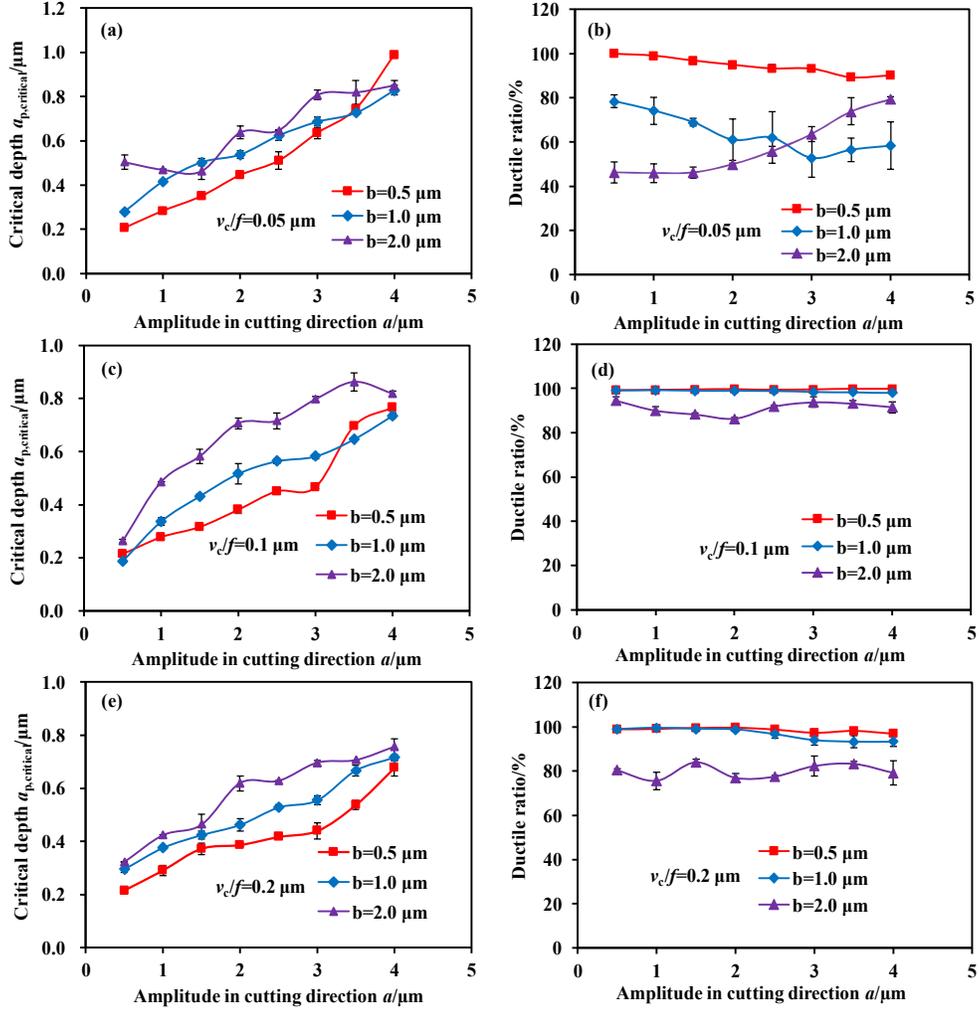

**Fig. 11.** Effects of vibration amplitude $a$ in the cutting direction on the critical depth and ductile ratio when $b$=0.5, 1, 2 μm. (a)-(b):$v_c/f$ is 0.05 μm; (c)-(d):$v_c/f$ is 0.1 μm; (e)-(f):$v_c/f$ is 0.2 μm.

### 4.3 Effects of vibration amplitude in the cutting direction

The increase of vibration amplitude $a$ in the cutting direction will also first enhance ductile-regime cutting of silicon, and then deteriorate the cutting performance. The effects of $a$ on the critical depth and ductile ratio are illustrated in Fig. 11. Here all six sets shown in Fig. 11 are used to demonstrate the effect of vibration amplitude $a$.

On the one hand, with the initial increase of vibration amplitude $a$ in the cutting direction, the critical depth is improved. As shown in Fig. 11(c) to Fig. 11(f), when $b$= 0.5 μm and 1 μm, the increase of vibration amplitude $a$ from 0.5 μm to 3 μm can significantly improve the critical depth by about three times, while the ductile ratio is close to 100%. This phenomenon can be attributed to the combined effects of the decrease in maximum undeformed chip thickness $h_{max}$ and effective cutting direction angle $\theta$, as illustrated in Fig. 9(a) and Fig. 9(c). With the increase of $a$, $h_{max}$ decreases while the upper bound $h_{ductile}$ increases due to the change of $\theta$, resulting in the improvement of critical depth.

On the other hand, with the further increase of vibration amplitude $a$ in the cutting direction,

the machined surface quality deteriorates gradually. As shown in Fig. 11(d) and Fig. 11(e), when $b$ = 0.5 and 1 μm, the ductile ratio shows a downward trend with a further increase of $a$. This observation is because that increase of $a$ will further decrease $h_{max}$ to reach the lower bound of $h_c$, causing the deterioration of surface quality. Another indirect evidence for the above statement can be found in Fig. 11(a) and Fig. 11(b). Even if $v_c/f$ is 0.05 μm < 0.1 μm, the ductile ratio can still be close to 100% if $a$ is small (0.5 μm). This is related to the direct relationship between the vibration amplitude $a$ and the maximum undeformed chip thickness $h_{max}$.

### 4.4 Summary of interactive effects of process parameters and optimization

As the previous subsections demonstrate, the four key process parameters have interactive rather than mutually independent effects on the ductile-to-brittle transition behavior in silicon cutting with elliptical vibration. This is due to their synergic effects on the undeformed chip thickness and the angle of effective cutting direction. The nominal cutting velocity only affects the maximum chip thickness but not the effective cutting direction; the vibration amplitudes in the DOC and cutting directions influence both the undeformed chip thickness and the effective cutting direction. However, to enhance the ductile-regime cutting of silicon, the maximum chip thickness must be bounded by the minimum chip thickness and the upper critical ductile chip thickness, both of which are further affected by the effective cutting direction. Hence, only when all the four key process parameters are all adequately determined, the ductile-regime cutting of silicon can be enhanced through the introduction of elliptical vibration.

Based on the above analysis and experimental results, the optimum process parameters for elliptical vibration cutting of silicon (crystal plane of (100) and crystal orientation of [110]) to achieve the maximum critical depth with acceptable surface quality are obtained as follows. The nominal cutting velocity and vibration frequency satisfy that $v_c/f$ = 0.1 μm. The vibration amplitude in the DOC direction $b$ equals to 1 μm. The vibration amplitude in the cutting direction $a$ equals to 4 μm. With this set of process parameters, the critical depth of silicon cutting in the ductile-regime can reach up to 700 nm.

It is worth mentioning that the ductile-to-brittle transition behavior of silicon varies with different crystal orientations in ductile-mode diamond cutting. As reported by Shamoto et al. (2014), the crystal orientation also affects the critical depth in elliptical vibration cutting of silicon. The results derived in this study are based on the cutting direction of [110] in the crystal plane of (110). The major findings – (1) the existence of a lower bound of nominal cutting velocity and (2) the relation between the vibration amplitudes and critical DOC – are universal to all crystal orientations. The quantitative effects of crystal orientation on the lower bound of nominal cutting velocity are worthy of future investigation by following the experimental procedure and data processing methods proposed in this study.

## 5. Conclusions

The minimum chip thickness and time-varying effective cutting direction have not been well considered in previous studies of elliptical vibration cutting of silicon, which leads to some inconsistent and contradictory results regarding its ductile-regime cutting performance. This study is devoted to the experimental and theoretical investigation of ductile-to-brittle transition behavior in the elliptical vibration cutting of silicon, with synergic consideration of the undeformed chip thickness and time-varying effective cutting direction. Systematic experiments have been performed to cover a wide span of process parameters. The effects of process parameters such as the nominal cutting velocity and elliptical trajectory on the critical depth and the ductile ratio of the machined surface have been evaluated. The conclusions can be drawn as follows:

(1) Elliptical vibration cutting has two unique characteristics over the conventional cutting, namely the reduced maximum chip thickness and enlarged effective cutting direction angle. To enhance the ductile-regime cutting of silicon, the maximum chip thickness in elliptical vibration cutting must be bounded between the minimum chip thickness and the upper critical ductile chip thickness. However, the enlarged effective cutting direction angle narrows the practical boundary of maximum chip thickness. If process parameters are not correctly determined, elliptical vibration cutting shows worse ductile-regime cutting performance over conventional diamond cutting due to its complex cutting characteristics.

(2) The nominal cutting velocity in elliptical vibration cutting only affects the maximum chip thickness but not the effective cutting direction. There exists a lower bound of nominal cutting velocity to enhance the ductile-regime cutting of silicon, besides the well-known upper bound. If the nominal cutting velocity is smaller than the lower bound value, the machined surface will be severely damaged.

(3) The increases of vibration amplitudes in both the DOC and cutting directions first enhance but then deteriorate the cutting performance, due to their simultaneous influences on the effective cutting direction and undeformed chip thickness.

(4) The initial increase of vibration amplitude in the cutting direction reduces the undeformed chip thickness that accounts for the improvement of critical depth. Its further increase, however, reduces the undeformed chip thickness towards the minimum chip thickness, which violates the stable cutting condition.

(5) The initial increase of vibration amplitude in the DOC direction also improves the critical depth due to the sharp reduction of undeformed chip thickness. However, its further increase deteriorates the surface quality severely due to the reverse of friction direction on the tool rake.

## Acknowledgments

This research was supported by the start-up fund from McCormick School of Engineering, Northwestern University, Evanston, IL, USA; the Innovation and Technology Fund, Hong Kong, #ITS/076/1; and the Shun Hing Institute of Advanced Engineering, Chinese University of Hong Kong (# RNE-p4-17).

## References

Abdulkadir, L. N., Abou-El-Hossein, K., Jumare, A. I., Odedeyi, P. B., Liman, M. M., Olaniyan, T. A., 2018. Ultra-precision diamond turning of optical silicon-a review. Int. J. Adv. Manuf. Technol. 96(1), 173-208.

Arif, M., Rahman, M., and San, W. Y., 2012. Analytical model to determine the critical conditions for the modes of material removal in the milling process of brittle material. J. Mater. Process. Technol. 212(9), 1925-1933.

Dai, H., Du, H., Chen, J., and Chen, G., 2019. Influence of elliptical vibration on the behavior of silicon during nanocutting. Int. J. Adv. Manuf. Technol. 1-16.

Fang, F. Z., Wu, H., Zhou, W., and Hu, X. T., 2007. A study on mechanism of nano-cutting single crystal silicon. J. Mater. Process. Technol. 184(1-3), 407-410.

Fang, F. Z., and Venkatesh, V. C., 1998. Diamond Cutting of Silicon with Nanometric Finish. CIRP Ann. - Manuf. Technol. 47(1), 45-49.

Fang, F. Z., and Zhang, G. X., 2003. An experimental study of edge radius effect on cutting single crystal silicon. Int. J. Adv. Manuf. Technol. 22(9-10), 703-707.

Geng, D., Liu, Y., Shao, Z., Zhang, M., Jiang, X., and Zhang, D., 2020. Delamination formation and suppression during rotary ultrasonic elliptical machining of CFRP. Compos. Part B-Eng. 183, 107698.

Geng, D., Lu, Z., Yao, G., Liu, J., Li, Z., and Zhang, D. 2017. Cutting temperature and resulting influence on machining performance in rotary ultrasonic elliptical machining of thick CFRP. Int. J. Mach. Tool Manu. 123, 160-170.

Gu, W., Zhu, Z., Zhu, W. L., Lu, L., To, S., and Xiao, G., 2018. Identification of the critical depth-of-cut through a 2D image of the cutting region resulting from taper cutting of brittle materials. Meas. Sci. Technol. 29(5), 055003.

Huang, W., Yu, D., Zhang, X., Zhang, M., and Chen, D., 2018. Ductile-regime machining model for ultrasonic elliptical vibration cutting of brittle materials. J. Manuf. Process. 36, 68-76.

Liu, K., Li, X. P., Rahman, M., Neo, K. S., and Liu, X. D., 2007. A study of the effect of tool cutting edge radius on ductile cutting of silicon wafers. Int. J. Adv. Manuf. Technol. 32(7-8), 631-637.

Nath, C., Rahman, M., and Neo, K. S., 2011. Modeling of the effect of machining parameters on maximum thickness of cut in ultrasonic elliptical vibration cutting. J. Manuf. Sci. E.-T. ASME. 133(1), 011007.

Son, S. M., Lim, H. S., and Ahn, J. H., 2005. Effects of the friction coefficient on the minimum cutting thickness in micro cutting. Int. J. Mach. Tool Manu. 45(4-5), 529-535.

Suzuki, N., Masuda, S., Haritani, M., & Shamoto, E. (2004, October). Ultraprecision micromachining of brittle materials by applying ultrasonic elliptical vibration cutting. In: Micro-Nanomechatronics and Human Science, 2004 and The Fourth Symposium Micro-Nanomechatronics for Information-Based Society, 2004. (pp. 133-138). IEEE.

Shamoto, E., and Suzuki N., 2014. Ultrasonic vibration diamond cutting and ultrasonic elliptical vibration cutting. In: Hashmi, M.S.J., Comprehensive Materials Processing, vol. 11, Elsevier, Waltham, pp. 405-454.

Wang, J., Du, H., Yang, Y., and Guo, P., 2019. An ultrafast 2-D non-resonant cutting tool for texturing micro-structured surfaces. J. Manuf. Process. Under review.

Wang, J., Yang, Y., and Guo, P., 2018. Effects of vibration trajectory on ductile-to-brittle transition in vibration cutting of single crystal silicon using a non-resonant tool. Procedia CIRP. 71, 289-292.

Wang, W., Yao, P., Wang, J., Huang, C., Zhu, H., Zou, B., Liu, H., and Yan, J., 2016. Crack-free ductile mode grinding of fused silica under controllable dry grinding conditions. Int. J. Mach. Tools Manuf. 109, 126-136.

Yan, J., Asami, T., Harada, H., and Kuriyagawa, T., 2009. Fundamental investigation of subsurface damage in single crystalline silicon caused by diamond machining. Precis. Eng. 33(4), 378-386.

Yan, J., Yoshino, M., Kuriagawa, T., Shirakashi, T., Syoji, K., and Komanduri, R., 2001. On the ductile machining of silicon for micro electro-mechanical systems (MEMS), opto-electronic and optical applications. Mat. Sci. Eng. A.-Strut. 297(1-2), 230-234.

Yuan, S., Zhang, C., Hu, J. 2015. Effects of cutting parameters on ductile material removal mode percentage in rotary ultrasonic face machining. P. I. Mech. Eng. B-J. Eng. Manu. 229(9), 1547-1556.

Yuan, Z. J., Zhou, M., and Dong, S., 1996. Effect of diamond tool sharpness on minimum cutting thickness and cutting surface integrity in ultraprecision machining. J. Mater. Process. Technol. 62(4), 327-330.

Zhang, J., Han, L., Zhang, J., Liu, H., Yan, Y., and Sun, T., 2019. Brittle-to-ductile transition in elliptical vibration-assisted diamond cutting of reaction-bonded silicon carbide. J. Manuf. Process. Accepted.

Zhang, J., Suzuki, N., Wang, Y., and Shamoto, E., 2014. Fundamental investigation of ultra-precision ductile machining of tungsten carbide by applying elliptical vibration cutting with single crystal diamond. J. Mater. Process. Technol. 214(11), 2644-2659.

Zhang, J., Zhang, J., Cui, T., Hao, Z., and Al Zahrani, A., 2017. Sculpturing of single crystal silicon microstructures by elliptical vibration cutting. J. Manuf. Process. 29(Supplement C), 389-398.

Zhang, X., Kumar, A. S., Rahman, M., Nath, C., and Liu, K., 2012. An analytical force model for orthogonal elliptical vibration cutting technique. J. Manuf. Process. 14(3), 378-387.

Zhu, Z., To, S., Xiao, G., Ehmann, K. F., and Zhang, G., 2016. Rotary spatial vibration-assisted diamond cutting of brittle materials. Precis. Eng. 44, 211-219.